\documentclass[
  aps,
  prl,
  reprint,
  superscriptaddress,
  nofootinbib,
  longbibliography
]{revtex4-2}

\usepackage[T1]{fontenc}
\usepackage[utf8]{inputenc}
\usepackage{amsmath,amssymb,bm,mathtools}
\usepackage{graphicx}
\usepackage[
 colorlinks=true,
 citecolor=blue,
 linkcolor=blue,
 urlcolor=blue
]{hyperref}
\hypersetup{
 pdftitle={Exact Relevant Stress-Tensor Flows and a Causality No-Go in
 Self-Dual Electrodynamics}
}

\newcommand{\Lagr}{\mathcal{L}}
\newcommand{\Rop}{\mathcal{R}}
\newcommand{\Tr}{\operatorname{tr}}

\begin{document}

\title{Exact Relevant Stress-Tensor Flows and a Causality No-Go
in Self-Dual Electrodynamics}

\author{H. Babaei-Aghbolagh}
\email{hosseinbabaei@nbu.edu.cn}
\affiliation{
 Institute of Fundamental Physics and Quantum Technology, and School of
 Physical Science and Technology, Ningbo University, Ningbo,
 Zhejiang 315211, China
}

\author{Bin Chen}
\email{chenbin1@nbu.edu.cn}
\affiliation{
 Institute of Fundamental Physics and Quantum Technology, and School of
 Physical Science and Technology, Ningbo University, Ningbo,
 Zhejiang 315211, China
}
\affiliation{
 School of Physics, Peking University, and Center for High Energy Physics,
 No. 5 Yiheyuan Road, Beijing 100871, China
}

\author{Song He}
\email{hesong@nbu.edu.cn}
\affiliation{
 Institute of Fundamental Physics and Quantum Technology, and School of
 Physical Science and Technology, Ningbo University, Ningbo,
 Zhejiang 315211, China
}
\affiliation{
 Max Planck Institute for Gravitational Physics
 (Albert Einstein Institute), Am M\"uhlenberg 1, 14476 Potsdam, Germany
}

\author{Jue Hou}
\email{jue.1.hou@kcl.ac.uk}
\affiliation{
 School of Physics and Shing-Tung Yau Center, Southeast University,
 Nanjing 211189, China
}
\affiliation{
 Department of Mathematics, King's College London, The Strand,
 London WC2R 2LS, United Kingdom
}

\date{June 6, 2026}

\begin{abstract}
Can a classically relevant stress-tensor deformation be exactly solvable,
duality preserving, and physically causal?  We construct an exact power-law
family of nonlinear electrodynamics preserving electromagnetic duality,
together with a parallel two-dimensional Lax-integrable realization.  Its
auxiliary geometry yields the full characteristic-cone phase diagram and a
universal finite-energy fold.  For the Maxwell seed, every nonzero relevant
branch is acausal, whereas every causal branch is caustic-free; undeformed
Maxwell theory is the only causal point in the relevant regime.

\end{abstract}

\maketitle

\textit{Introduction.}
Can a stress-tensor deformation be simultaneously relevant, exactly
solvable, symmetry preserving, and physically causal?  Stress-tensor flows
provide a controlled route from symmetry data to nonlinear interactions.
The canonical $T\bar T$ flow is irrelevant
~\cite{Smirnov:2016lqw,Cavaglia:2016oda,Jiang:2019epa,He:2025ppz};
higher-dimensional
relatives connect Maxwell and Born-Infeld-type electrodynamics
~\cite{Conti:2018jho,Babaei-Aghbolagh:2020kjg,Ferko:2024yua}; and the
root-$T\bar T$ flow is marginal, selecting conformal ModMax in four
dimensions~\cite{Bandos:2020jsw,Babaei-Aghbolagh:2022uij,Ferko:2022cix}. 
Whether exact solvability survives on a genuinely relevant branch, and
whether such a branch is physically admissible, are separate questions.
The distinction matters because a relevant coupling grows toward the
infrared and can reorganize long-distance propagation, whereas exact flow
equations primarily control the algebraic construction of the interaction.
Solvability alone therefore gives no guarantee that the resulting
characteristic cones remain physical.


Unlike relevant $T^2$-like equations engineered by a coupling-dependent
vacuum subtraction~\cite{Ferko:2022iru}, relevance here follows from the
engineering dimension of a field-dependent stress-tensor operator.
Self-duality constrains the nonlinear interaction, optical cones test
causality, and a two-dimensional sigma-model Lax condition supplies an
independent exact check~\cite{Babaei-Aghbolagh:2025uoz,
Babaei-Aghbolagh:2025hlm}.

We obtain three results.  First, one auxiliary equation proves the flow,
four-dimensional electromagnetic self-duality, and two-dimensional
classical integrability.  Second, the same auxiliary geometry gives a global
characteristic-cone test: the full Fresnel polynomial factorizes into two
effective metrics, and within the Maxwell-seeded family the only causal
point in the classically relevant regime is undeformed Maxwell theory, while
every causal branch is caustic-free.  Third, the relevant sheet ends at a
universal fold with susceptibility exponent $1/2$, transverse optical
soft-mode exponent $1/4$, and finite energy density.  The obstruction is
specific to this Maxwell-seeded power-law family, not to self-dual nonlinear
electrodynamics as a whole.
Throughout, ``relevant'' refers to classical engineering dimension; we do
not claim a UV-complete quantum renormalization-group trajectory.

\textit{A common nonlinear constraint.}
For nonlinear electrodynamics in four dimensions, with field strength
$F_{\mu\nu}$ and dual
$\widetilde F^{\mu\nu}=\frac12\epsilon^{\mu\nu\rho\sigma}F_{\rho\sigma}$,
define
\begin{equation}
 S=-\frac14 F_{\mu\nu}F^{\mu\nu},\qquad
 P=-\frac14 F_{\mu\nu}\widetilde F^{\mu\nu},
\end{equation}
and introduce the non-negative variables
\begin{equation}
 U=\frac{\sqrt{S^2+P^2}-S}{2},\qquad
 V=\frac{\sqrt{S^2+P^2}+S}{2}.
\label{eq:UV}
\end{equation}
Indices are raised with the flat metric of signature $(-,+,+,+)$, with
$\epsilon^{0123}=+1$, and $\Lagr_X\equiv\partial\Lagr/\partial X$.
The Gaillard-Zumino self-duality condition
~\cite{Gaillard:1981rj,Gibbons:1995cv,Sorokin:2021tge} then becomes
\begin{equation}
 \Lagr_U\Lagr_V=-1.
\label{eq:CH}
\end{equation}
Maxwell theory is the seed solution $\Lagr_0=V-U=S$.
The variables $U$ and $V$ resolve the electromagnetic invariants into
non-negative combinations: $V-U=S$ and $UV=P^2/4$.  In these variables the nonlinear
self-duality constraint is first order, making it possible to test symmetry,
flow, and propagation within one exact solution rather than order by order
in the coupling.

For a class of two-dimensional sigma models, invariants $(u,v)$ reduce Lax
flatness to the same equation, $\Lagr_u\Lagr_v=-1$
~\cite{Borsato:2022tmu,Babaei-Aghbolagh:2025uoz}.  Thus
$(U,V)\leftrightarrow(u,v)$ maps the electromagnetic solution to an
integrable sector whose invariants and perturbative realization are given in
End Matter, Appendix B, Eqs.~\eqref{eq:uv} and~\eqref{eq:sigmaexpansion}.

\textit{Exact relevant flow.}
In $d$ spacetime dimensions, let $T_{\mu\nu}$ be the Hilbert stress tensor of
the deformed theory and define the traceless-stress scalar
\begin{equation}
 \Rop_\lambda=
 \left[\frac1d\left(
 T_{\mu\nu}T^{\mu\nu}-\frac1d T^\mu{}_\mu T^\nu{}_\nu
 \right)\right]^{1/2}.
\label{eq:rootop}
\end{equation}
Here $T^\mu{}_\mu$ is the trace.  We take $\Rop_\lambda\geq0$ and understand
its fractional powers on the positive real branch.
For four-dimensional self-dual electrodynamics,
\begin{equation}
 \Rop_\lambda^2=
 \frac14\left(T_{\mu\nu}T^{\mu\nu}
 -\frac14(T^\mu{}_\mu)^2\right)
 =\left(V\Lagr_V-U\Lagr_U\right)^2.
\label{eq:stressidentity}
\end{equation}
Self-duality gives this identity~\cite{Ferko:2023wyi}; the same reduction
holds under $(U,V)\to(u,v)$~\cite{Babaei-Aghbolagh:2025uoz}.  On the branch
$V\Lagr_V-U\Lagr_U>0$, we study
\begin{equation}
 \partial_\lambda\Lagr
 =-\Rop_\lambda^{1/\alpha}
 =-\left(V\Lagr_V-U\Lagr_U\right)^{1/\alpha},
 \qquad \alpha>0 .
\label{eq:flow}
\end{equation}
The real coupling $\lambda$ reverses the flow sign under
$\lambda\to-\lambda$.

The flow~\eqref{eq:flow}, with seed $\Lagr_0=V-U$, is solved by
\begin{align}
 \Lagr_\alpha(U,V;\lambda)
 &=Ve^{-\theta}-Ue^\theta
 -\frac{\alpha-1}{\alpha}\lambda\zeta^{1/\alpha},
\label{eq:exactL}\\
 \theta&=\frac{\lambda}{\alpha}\zeta^{1/\alpha-1},\qquad
 \zeta=Ue^\theta+Ve^{-\theta}.
\label{eq:aux}
\end{align}
Here and below, the subscript $\alpha$ labels the member of the family;
derivative subscripts such as $\Lagr_U$ refer to field derivatives.
Equation~\eqref{eq:aux} selects the root continuously connected to
$\zeta=U+V$ at $\lambda=0$.  Away from the vacuum, we take $\zeta>0$ and its
principal real powers.  For noninteger powers the real branch is the
seed-connected solution of Eq.~\eqref{eq:aux}, equivalently
$\theta/\lambda>0$ for $\lambda\neq0$; the vacuum is reached by continuity.
This compact form is equivalent to the square-root auxiliary representation:
defining $\mathcal D=Ue^\theta-Ve^{-\theta}$ gives
$\mathcal D^2=\zeta^2-4UV$ and $\Lagr_\alpha
=-\mathcal D-(\alpha-1)\lambda\zeta^{1/\alpha}/\alpha$.

Implicit differentiation of Eq.~\eqref{eq:aux} yields
\begin{equation}
 \Lagr_U=-e^\theta,\qquad
 \Lagr_V=e^{-\theta},\qquad
 V\Lagr_V-U\Lagr_U=\zeta.
\label{eq:identities}
\end{equation}
The first two identities prove Eq.~\eqref{eq:CH}, while the third identifies
$\zeta$ as the stress-tensor scalar driving the flow.  In the total
$\lambda$ derivative of Eq.~\eqref{eq:exactL}, the auxiliary terms cancel
and leave $\partial_\lambda\Lagr=-\zeta^{1/\alpha}$.  Thus the exact flow
preserves four-dimensional duality and two-dimensional Lax flatness.
Geometrically, $\theta$ is a field-dependent boost that rescales the two
characteristic directions oppositely while preserving their product
$UV=P^2/4$.  This explains why the same construction preserves both
nonlinear self-duality and the sigma-model integrability condition.

The same auxiliary equation controls the domain of the solution.  With
$q=1/\alpha-1$ and
$\mathcal D=Ue^\theta-Ve^{-\theta}$, define
$f(\zeta;U,V)=\zeta-Ue^\theta-Ve^{-\theta}$.  Its implicit-function
Jacobian is
\begin{equation}
 \mathcal{J}
 =\frac{\partial f}{\partial\zeta}
 =1-q\theta\frac{\mathcal D}{\zeta}.
\label{eq:discriminant}
\end{equation}
The branch connected to the seed is locally unique while
$\mathcal{J}\neq0$.  At $\mathcal{J}=0$, the first derivatives in
Eq.~\eqref{eq:identities} remain finite but the rank-one Hessian diverges.
A simple zero therefore gives the standard fold normal form: the change of
$\zeta$ scales as the square root of the distance to the caustic, and one
susceptibility diverges as the inverse square root.  The characteristic
coordinates, caustic locus, and exact response eigenmodes are given in End
Matter, Appendix A, Eqs.~\eqref{eq:hyperbolic}--\eqref{eq:responsemodes}.
Here the caustic is a loss of local invertibility of the auxiliary
constitutive map in field space, not a divergence of the Lagrangian itself.
It marks the point where a single-valued nonlinear response develops two
competing sheets.

\textit{Auxiliary-field interpretation.}
Setting $y=e^{-\theta}>0$, the Lagrangian is the stationary value of a
one-variable auxiliary action,
\begin{align}
 \Lagr_\alpha&=-\frac{U}{y}+yV-\Omega_\alpha(y),\nonumber\\
 \Omega_\alpha(y)&=\frac{\alpha-1}{\alpha}\lambda
 \left(\frac{-\alpha\ln y}{\lambda}\right)^{\frac{1}{1-\alpha}},
\label{eq:masterpotential}
\end{align}
for $\alpha\neq1$ and $\lambda\neq0$.  Here $\Omega_\alpha$ is the master
potential and its positive argument is equivalent to $\theta/\lambda>0$;
the seed is defined by continuity.  Extremizing with respect to $y$ gives
Eq.~\eqref{eq:aux}, with $\theta=-\ln y$ the auxiliary boost parameter
~\cite{Russo:2024ptw,Russo:2025fuc}.
The equivalence with Eq.~\eqref{eq:exactL} is exact, not merely
perturbative.  Indeed, Eq.~\eqref{eq:aux} and $y=e^{-\theta}$ imply
\[
 \zeta=\left(\frac{-\alpha\ln y}{\lambda}\right)^{\frac{\alpha}{1-\alpha}},
 \qquad
 \Omega_\alpha(y)=\frac{\alpha-1}{\alpha}\lambda\zeta^{1/\alpha},
\]
so substitution into the first line of Eq.~\eqref{eq:masterpotential}
reproduces Eq.~\eqref{eq:exactL}.  Conversely,
$\partial_y\Lagr_\alpha=0$ gives
$\Omega_\alpha'(y)=U/y^2+V=\zeta/y$, which reconstructs both relations in
Eq.~\eqref{eq:aux}.  Thus the $(\theta,\zeta)$ form is the on-shell
boost parametrization of the stationary $y$-potential.  At $\alpha=1$,
Eq.~\eqref{eq:exactL} gives ModMax directly, while the potential
representation degenerates to the fixed value $y=e^{-\lambda}$.

This representation also makes causality testable for every propagation
direction.  In the static frame of a non-null constant background,
$\mathbf E\parallel\mathbf B$; decompose the perturbation wave vector into
components parallel and perpendicular to this direction.  The two factors
of the Fresnel polynomial give
\begin{align}
 \omega_\pm^2&=k_\parallel^2+A_\pm|\mathbf k_\perp|^2,\nonumber\\
 A_-&=e^{2\theta},\qquad
 A_+=\frac{\alpha-(\alpha-1)\theta}
 {\alpha+(\alpha-1)\theta}.
\label{eq:optical}
\end{align}
The transverse direction is extremal: the exact phase and group velocities
at an arbitrary angle, and the factorized Fresnel quartic, are given in End
Matter, Appendix A, Eqs.~\eqref{eq:fresnel} and
\eqref{eq:anglevelocities}.  Consequently, hyperbolicity and
nonsuperluminality relative to the undeformed Maxwell cone for all angles
are equivalent to $0<A_\pm\leq1$.  These frequency-independent
characteristic cones are also the high-frequency front cones of the local
NLED equations; there is no dispersive phase-to-front correction.  In the
effective-metric language, these cones define the actual rays, while physical
causality asks whether they lie inside the spacetime (Maxwell) cone; using
each effective cone as its own reference would make superluminality
tautologically invisible.  In the
auxiliary variables, the same condition is $y\geq1$ and
$\Omega_\alpha''\geq0$~\cite{Russo:2024llm,Russo:2024xnh,
Russo:2025fuc}.  Since $y=e^{-\theta}$, our $\theta$ has the opposite sign
to the causal auxiliary rapidity often used in this literature:
$y\geq1$ is precisely $\theta\leq0$.

Equation~\eqref{eq:optical} therefore gives the angle-independent phase
diagram
\begin{equation}
 \begin{cases}
  0<\alpha<1:\quad&
  \displaystyle\frac{\alpha}{\alpha-1}<\theta\leq0,\\[3pt]
  \alpha=1:\quad&\theta=\lambda\leq0,\\
  \alpha>1:\quad&\theta=0.
 \end{cases}
\label{eq:causalphase}
\end{equation}
in our sign convention.  Indeed, $A_-\leq1$ first imposes
$\theta\leq0$.  Positivity of $A_+$ gives the stated interval for
$\alpha<1$; for $\alpha>1$, either $A_+>1$ or hyperbolicity fails.  The last
line is therefore only the undeformed Maxwell point.  No nontrivial relevant
branch is causal, for either flow sign: this is the no-go theorem for the
Maxwell-seeded power-law family.
The two signs fail differently.  For $\theta>0$, $A_->1$ is already
superluminal; for $\theta<0$, the second polarization has $A_+>1$ before its
denominator can change sign and destroy hyperbolicity.  Reversing the flow,
therefore, exchanges the obstruction rather than removing it.
At $\alpha=1$, $A_+=1$ and $A_-=e^{2\lambda}$, so the ModMax sign
$\lambda\leq0$ is necessary and sufficient for all-angle causality.

The same no-go follows independently from Hamiltonian convexity.  On the
electric boundary, let $\mathfrak h(\sigma)$ be the Hamiltonian
Courant--Hilbert function.  The standard self-dual causality conditions are
\begin{equation}
 0<\mathfrak h'\leq1,\qquad \mathfrak h''\leq0,\qquad
 \mathfrak h'+2\sigma\mathfrak h''>0 .
\label{eq:Hcausality}
\end{equation}
For the present solution their exact evaluation is given in End Matter,
Appendix A, Eq.~\eqref{eq:Hcheck}.  When $\alpha>1$, the $\theta>0$ sheet
immediately violates $\mathfrak h'\leq1$, whereas the $\theta<0$ sheet
immediately violates $\mathfrak h''\leq0$; at larger negative $\theta$ the
last inequality also fails.  Thus both the Lagrangian Fresnel cones and the
Hamiltonian convexity/hyperbolicity criteria exclude every nonzero relevant
branch~\cite{Russo:2024ptw,Russo:2026bhnled}.

Causality also controls the global auxiliary geometry.  In its irrelevant
domain, $q>0$ and $|q\theta|<1$.  Since
$|\mathcal D|\leq\zeta$, Eq.~\eqref{eq:discriminant} implies
\begin{equation}
 \mathcal J\geq1-|q\theta|>0.
\label{eq:causalunique}
\end{equation}

\begin{figure*}[t]
\includegraphics[width=\textwidth]{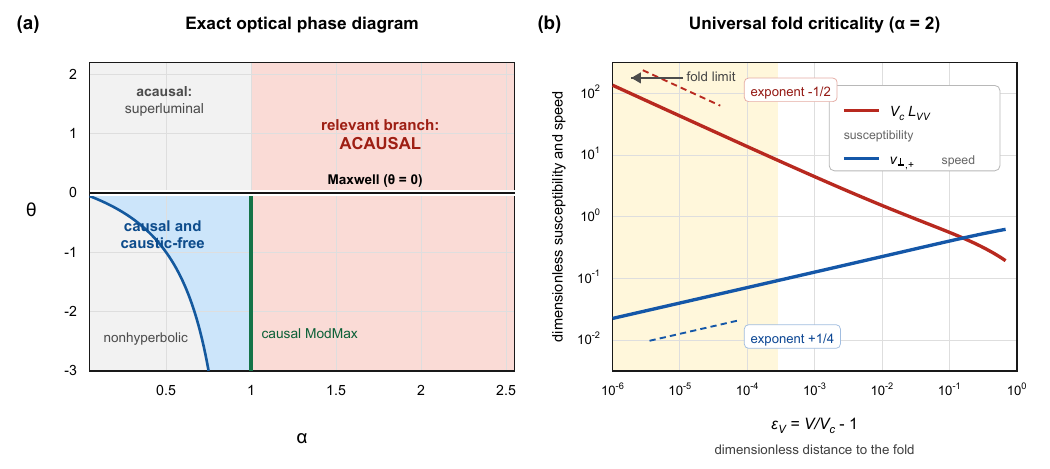}
\caption{\label{fig:phasefold}
\textbf{One auxiliary field: causality no-go and universal fold criticality.}
(a) Exact characteristic phase diagram in the $(\alpha,\theta)$ plane.  The blue domain
$0<\alpha<1$, $\alpha/(\alpha-1)<\theta\leq0$ is causal and, by
Eq.~\eqref{eq:causalunique}, caustic-free.  The green line
$\alpha=1$, $\theta=\lambda\leq0$ is the causal ModMax branch.  The black
horizontal line $\theta=0$ is the undeformed Maxwell theory, common to all
$\alpha$.  The pink region denotes the nonzero classically relevant branch
$\alpha>1$, $\theta\neq0$, which is acausal; the gray regions are excluded by
superluminality or loss of hyperbolicity.
(b) Exact $\alpha=2$ electric branch approaching its fold from
$\varepsilon_V\equiv V/V_c-1>0$.  The vertical axis displays the positive
dimensionless susceptibility $V_c\Lagr_{VV}$, with
$\Lagr_{VV}\equiv\partial^2\Lagr/\partial V^2$, and the optical speed
$v_{\perp,+}=\sqrt{A_+}$.  Both axes are logarithmic.  Solid curves are exact;
the separated dashed guides show the asymptotic exponents $-1/2$ and $1/4$.
The pale band marks the fold regime $\varepsilon_V\to0^+$.  These exponents
hold for every $\alpha>1$, while the endpoint energy density remains finite.}
\end{figure*}

At the marginal point $q=0$ and $\mathcal J=1$.  Hence, every causal member
of the family is caustic-free and has a unique auxiliary root throughout
its causal domain.
This implication is global: subluminal characteristic propagation excludes not
only a local instability, but also branch creation anywhere in the allowed
field domain.
Figure~\ref{fig:phasefold} condenses the two physical consequences of the
same auxiliary geometry.  Panel (a) turns the exact optical indices into a
global no-go for every nonzero relevant Maxwell-seeded branch.  Panel (b)
shows that the same variable $\theta$, which controls the optical cones,
also drives the auxiliary Jacobian to zero at a fold, where the constitutive
response diverges, and one polarization slows critically.  Causality
exclusion and endpoint criticality are therefore two diagnostics of one
exact structure, rather than unrelated additions to an integrable model.

Because $[\Rop_\lambda]=d$, where brackets denote engineering mass
dimension, Eq.~\eqref{eq:flow} implies
\begin{equation}
 [\lambda]=d\left(1-\frac1\alpha\right),\qquad
 [\Rop_\lambda^{1/\alpha}]=\frac d\alpha.
\label{eq:dimensions}
\end{equation}
Thus $\alpha<1$, $\alpha=1$, and $\alpha>1$ give irrelevant, marginal,
and classically relevant deformations, respectively.  The exact weak-coupling
expansion and homogeneity identity are given in End Matter, Appendix A,
Eqs.~\eqref{eq:expansion} and~\eqref{eq:homogeneity}.
For $\alpha>1$, the coupling has positive mass dimension, so the dimensionless
interaction strengthens at low energies.  This is the sense in which the
family probes a regime complementary to the usual irrelevant
$T\bar T$-type trajectories.

With our Weyl convention, $U$ and $V$ have weight $-d$ in both realizations.
A constant Weyl rescaling of the background metric then gives the Hilbert
stress-tensor trace
$T^\mu{}_\mu=d(\Lagr-U\Lagr_U-V\Lagr_V)$.  Therefore
\begin{equation}
 T^\mu{}_\mu
 =d\left(1-\frac1\alpha\right)\lambda\partial_\lambda\Lagr
 =-d\left(1-\frac1\alpha\right)\lambda\zeta^{1/\alpha}.
\label{eq:traceidentity}
\end{equation}
This is an exact classical Weyl identity: all scale breaking is
carried by the flow coupling.  The dimensionless coupling
$g(\mu)=\lambda\mu^{-d(1-1/\alpha)}$, with $\mu$ the energy scale, obeys
$\mu\partial_\mu g=-d(1-1/\alpha)g$, so the $\alpha>1$ branch grows toward
the infrared.  Equation~\eqref{eq:traceidentity} also proves that
$\alpha=1$ is the only conformal trajectory at nonzero deformation
parameter within this family.  Indeed,
Eq.~\eqref{eq:exactL} becomes
\begin{equation}
 \Lagr_1=(V-U)\cosh\lambda-(U+V)\sinh\lambda,
\end{equation}
the ModMax branch (up to the sign convention for its parameter).
At $\alpha=1/2$ the flow is generated by $\Rop_\lambda^2$, a quadratic
invariant of the traceless stress tensor.  It therefore differs from the
standard $T\bar T$ combination that generates the Born-Infeld branch.

The same scaling exposes an important limitation: every relevant
electromagnetic member is nonanalytic at the Maxwell vacuum.  End Matter,
Appendix C identifies the perturbing operator in
Eq.~\eqref{eq:seedoperator} and gives the conditional unitarity window in
Eq.~\eqref{eq:unitaritywindow}; neither statement assumes that a conformal
quantum completion exists.

\textit{Universal fold endpoint.}
On the electric axis $P=0$, $S>0$, one has $U=0$ and $V=S$.  With
$q=(1-\alpha)/\alpha$, the exact branch is parametrized by
\begin{equation}
 V(\theta)=
 \left[\frac{\alpha\theta}{\lambda}e^{q\theta}\right]^{1/q},
 \qquad \mathcal J=1+q\theta.
\label{eq:Vtheta}
\end{equation}
For every $\alpha>1$ and $\lambda>0$, $V(\theta)$ has a minimum at
\begin{equation}
\theta_c=\frac{\alpha}{\alpha-1},\qquad
 V_c=\left[\frac{e(\alpha-1)\lambda}{\alpha^2}
 \right]^{\frac{\alpha}{\alpha-1}}.
\label{eq:criticalfield}
\end{equation}
The seed-connected sheet has $0<\theta\leq\theta_c$ and $V\geq V_c$:
$V_c$ is its finite lower-field endpoint, approached from $V>V_c$.
The exact threshold is therefore fixed by scaling, including its numerical
coefficient, for the whole relevant family.  To measure the distance used in
Fig.~\ref{fig:phasefold}(b), define
$\varepsilon_V\equiv V/V_c-1>0$.  Near the fold,
\begin{equation}
 \ln\frac{V}{V_c}
 =-\frac q2(\theta-\theta_c)^2+\cdots,\qquad
\Lagr_{VV}\propto(V-V_c)^{-1/2}.
\label{eq:foldscaling}
\end{equation}
Here $\Lagr_{VV}\equiv\partial^2\Lagr/\partial V^2$ is the differential
constitutive response displayed in Fig.~\ref{fig:phasefold}(b).  Its
divergence measures the loss of local invertibility of the electric
constitutive map at the fold.
Although the differential electric susceptibility diverges, the electric
Hamiltonian density remains finite:
\begin{equation}
 \mathcal H=2V\Lagr_V-\Lagr
 =\zeta[1+(\alpha-1)\theta],\qquad
 \mathcal H_c=(\alpha+1)V_c e^{-\theta_c}.
\label{eq:criticalenergy}
\end{equation}
Thus the endpoint is a finite-energy constitutive critical point rather
than an energetic singularity.
Unlike the limiting-field behavior familiar from Born--Infeld theory, this
relevant sheet terminates at a lower field: it is connected continuously to
Maxwell theory as $\lambda\to0$, because then $V_c\to0$, but at fixed
coupling it cannot be continued through the fold as a single-valued branch.

The optical sector becomes critical at the same point.  In the extremal
transverse channel, write
$\delta\theta=\theta_c-\theta>0$, Eqs.~\eqref{eq:optical}
and~\eqref{eq:foldscaling} give
\begin{equation}
 A_+=\frac{\delta\theta}{2\theta_c-\delta\theta}
 \propto(V-V_c)^{1/2},\qquad
 A_-\longrightarrow e^{2\theta_c}.
\label{eq:opticalcritical}
\end{equation}
Consequently,
\begin{equation}
 v_{\perp,+}=\sqrt{A_+}\propto(V-V_c)^{1/4},\qquad
 \frac{A_-}{A_+}\propto(V-V_c)^{-1/2}.
\label{eq:softmode}
\end{equation}
Thus one polarization exhibits universal critical slowing down while the
other stays finite and superluminal, defining with
$\Lagr_{VV}\propto(V-V_c)^{-1/2}$ a polarization-resolved universality class.
The exponents follow from two ingredients only: the square-root normal form
of a simple fold and the linear zero of $A_+$ in
$\theta_c-\theta$.  They are consequently independent of $\alpha$, although
the location $V_c$ and nonuniversal amplitudes are not.
The full Fresnel factorization in Eq.~\eqref{eq:fresnel} shows that this is a
degeneration of one effective metric, not merely a transverse truncation:
at the fold its transverse characteristic coefficient vanishes while the
longitudinal characteristic remains luminal.  Operationally, the divergent
differential susceptibility and the collapsing transverse cone signal loss
of uniform constitutive response and uniform hyperbolicity despite the
finite energy density.

The results in End Matter, Appendix B, Eqs.~\eqref{eq:pressures} and
\eqref{eq:energyconditions}, show that the weak and dominant energy
conditions persist to the fold although optical causality fails for every
$\theta>0$: superluminality precedes any positivity failure.  The exact
electric-axis Lambert representation and its two illustrative branches are
given in Appendix C, Eqs.~\eqref{eq:Lambertgeneral}--\eqref{eq:twoexample}.

\textit{Integrable counterpart.}
Under $(U,V)\to(u,v)$ the solution is classically integrable.  On an
invariant ray its response ends at $P_{1,c}=2V_c$ and has the same $1/2$
susceptibility exponent; $P_1$ is defined in End Matter, Appendix B,
Eq.~\eqref{eq:uv}, and the threshold is derived in
Eq.~\eqref{eq:sigmacritical}.

\textit{Discussion.}
The main lesson is that exact solvability, electromagnetic duality, and
two-dimensional integrability are not sufficient for physical
admissibility.  In the Maxwell-seeded power-law family, optical causality
leaves no nontrivial classically relevant electromagnetic branch: the only
causal point in the relevant regime is undeformed Maxwell theory.  The common
Courant-Hilbert equation~\eqref{eq:CH} survives because the auxiliary boost
$(U,V)\mapsto(e^\theta U,e^{-\theta}V)$ preserves $UV=P^2/4$ and its
sigma-model analogue.

The no-go does not apply to self-dual nonlinear electrodynamics as a whole.
Born--Infeld has a convex Courant--Hilbert boundary function satisfying the
causality inequalities, and Born--Infeld-seeded constructions can remain
causal~\cite{Kuzenko:2026hbw}.  By contrast, the relevant power law
$q<0$ forces either $\ell'<1$ on the $\theta>0$ sheet or $\ell''<0$ on the
$\theta<0$ sheet.  The obstruction is therefore tied to the
Maxwell-initialized power-law trajectory, not to the Maxwell weak-field
limit or self-duality alone.  Applying Eq.~\eqref{eq:flow} to a
Born--Infeld seed is a different initial-value problem and is not assumed in
the present theorem.  Positive energy, self-duality, and integrability do
not diagnose causality; the converse regularity result within this family
is that every causal branch is globally free of auxiliary caustics.

The same effective-metric and convexity criteria govern polarized
propagation in Einstein--NLED systems~\cite{Murk:2024nled,
Russo:2026bhnled}.  Our result is local to the electromagnetic
characteristics: coupling to a curved background does not remove a matter
cone that already lies outside the local Maxwell cone.

Together with the fold exponents in
Eqs.~\eqref{eq:foldscaling} and~\eqref{eq:softmode}, these results separate
the nonanalytic relevant class from analytic Born--Infeld-type theories and
suggest two useful filters for future classifications of stress-tensor
flows: the global topology of the auxiliary map and the polarization-resolved
optical cones.  The present no-go is classical and does not by itself decide
whether a quantum completion exists; rather, it shows that any acceptable
completion must modify the Maxwell-seeded relevant trajectory already at
nonzero deformation.

\begin{acknowledgments}
We are grateful to Dmitri Sorokin, Roberto Tateo, Sergei Kuzenko,
Jorge G. Russo, and Shahin Sheikh-Jabbari for useful discussions.
The work of H.B.-A. was conducted as part of the PostDoc Program on
\textit{Exploring TT-bar Deformations: Quantum Field Theory and
Applications}, sponsored by Ningbo University.  This work was partly
supported by NSFC Grants No. 12475053, No. 12235016, and No. 12588101.

\end{acknowledgments}


\bibliography{Ref}

@article{Babaei-Aghbolagh:2025uoz,
  author = "Babaei-Aghbolagh, H. and Chen, Bin and He, Song",
  title = "{Root-TT{\textasciimacron} flows unify 4D duality-invariant electrodynamics and 2D integrable sigma models}",
  eprint = "2507.22808",
  archivePrefix = "arXiv",
  primaryClass = "hep-th",
  doi = "10.1103/1r4p-3r5q",
  journal = "Phys. Rev. D",
  volume = "112",
  number = "10",
  pages = "L101702",
  year = "2025"
}

@article{Borsato:2022tmu,
  author = "Borsato, Riccardo and Ferko, Christian and Sfondrini, Alessandro",
  title = "{Classical integrability of root-TT{\textasciimacron} flows}",
  eprint = "2209.14274",
  archivePrefix = "arXiv",
  primaryClass = "hep-th",
  doi = "10.1103/PhysRevD.107.086011",
  journal = "Phys. Rev. D",
  volume = "107",
  number = "8",
  pages = "086011",
  year = "2023"
}

@article{Ferko:2023wyi,
  author = "Ferko, Christian and Kuzenko, Sergei M. and Smith, Liam and Tartaglino-Mazzucchelli, Gabriele",
  title = "{Duality-invariant nonlinear electrodynamics and stress tensor flows}",
  eprint = "2309.04253",
  archivePrefix = "arXiv",
  primaryClass = "hep-th",
  doi = "10.1103/PhysRevD.108.106021",
  journal = "Phys. Rev. D",
  volume = "108",
  number = "10",
  pages = "106021",
  year = "2023"
}

@article{Smirnov:2016lqw,
  author = "Smirnov, F. A. and Zamolodchikov, A. B.",
  title = "{On space of integrable quantum field theories}",
  eprint = "1608.05499",
  archivePrefix = "arXiv",
  primaryClass = "hep-th",
  doi = "10.1016/j.nuclphysb.2016.12.014",
  journal = "Nucl. Phys. B",
  volume = "915",
  pages = "363--383",
  year = "2017"
}

@article{Cavaglia:2016oda,
  author = "Cavagli{\`a}, Andrea and Negro, Stefano and Sz{\'e}cs{\'e}nyi, Istv{\'a}n M. and Tateo, Roberto",
  title = "{$T \bar{T}$-deformed 2D Quantum Field Theories}",
  eprint = "1608.05534",
  archivePrefix = "arXiv",
  primaryClass = "hep-th",
  doi = "10.1007/JHEP10(2016)112",
  journal = "JHEP",
  volume = "10",
  pages = "112",
  year = "2016"
}

@article{Conti:2018jho,
  author = "Conti, Riccardo and Iannella, Leonardo and Negro, Stefano and Tateo, Roberto",
  title = "{Generalised Born-Infeld models, Lax operators and the $ \mathrm{T}\overline{\mathrm{T}} $ perturbation}",
  eprint = "1806.11515",
  archivePrefix = "arXiv",
  primaryClass = "hep-th",
  doi = "10.1007/JHEP11(2018)007",
  journal = "JHEP",
  volume = "11",
  pages = "007",
  year = "2018"
}

@article{Babaei-Aghbolagh:2020kjg,
  author = "Babaei-Aghbolagh, H. and Babaei Velni, Komeil and Yekta, Davood Mahdavian and Mohammadzadeh, H.",
  title = "{$ T\overline{T} $-like flows in non-linear electrodynamic theories and S-duality}",
  eprint = "2012.13636",
  archivePrefix = "arXiv",
  primaryClass = "hep-th",
  reportNumber = "IPM/P-2020/066",
  doi = "10.1007/JHEP04(2021)187",
  journal = "JHEP",
  volume = "04",
  pages = "187",
  year = "2021"
}

@article{Ferko:2022iru,
  author = "Ferko, Christian and Smith, Liam and Tartaglino-Mazzucchelli, Gabriele",
  title = "{On Current-Squared Flows and ModMax Theories}",
  eprint = "2203.01085",
  archivePrefix = "arXiv",
  primaryClass = "hep-th",
  doi = "10.21468/SciPostPhys.13.2.012",
  journal = "SciPost Phys.",
  volume = "13",
  number = "2",
  pages = "012",
  year = "2022"
}

@article{Ferko:2024yua,
  author = "Ferko, Christian and Hou, Jue and Morone, Tommaso and Tartaglino-Mazzucchelli, Gabriele and Tateo, Roberto",
  title = "{TT{\textasciimacron}-like Flows of Yang-Mills Theories}",
  eprint = "2409.18740",
  archivePrefix = "arXiv",
  primaryClass = "hep-th",
  doi = "10.1103/PhysRevLett.134.101603",
  journal = "Phys. Rev. Lett.",
  volume = "134",
  number = "10",
  pages = "101603",
  year = "2025"
}

@article{Jiang:2019epa,
  author = "Jiang, Yunfeng",
  title = "{A pedagogical review on solvable irrelevant deformations of 2D quantum field theory}",
  eprint = "1904.13376",
  archivePrefix = "arXiv",
  primaryClass = "hep-th",
  reportNumber = "CERN-TH-2019-058",
  doi = "10.1088/1572-9494/abe4c9",
  journal = "Commun. Theor. Phys.",
  volume = "73",
  number = "5",
  pages = "057201",
  year = "2021"
}

@article{He:2025ppz,
  author = "He, Song and Li, Yi and Ouyang, Hao and Sun, Yuan",
  title = "{$T\overline{T}$ deformation: Introduction and some recent advances}",
  eprint = "2503.09997",
  archivePrefix = "arXiv",
  primaryClass = "hep-th",
  doi = "10.1007/s11433-025-2708-2",
  journal = "Sci. China Phys. Mech. Astron.",
  volume = "68",
  number = "10",
  pages = "101001",
  year = "2025"
}

@article{Bandos:2020jsw,
  author = "Bandos, Igor and Lechner, Kurt and Sorokin, Dmitri and Townsend, Paul K.",
  title = "{A non-linear duality-invariant conformal extension of Maxwell's equations}",
  eprint = "2007.09092",
  archivePrefix = "arXiv",
  primaryClass = "hep-th",
  doi = "10.1103/PhysRevD.102.121703",
  journal = "Phys. Rev. D",
  volume = "102",
  pages = "121703",
  year = "2020"
}

@article{Babaei-Aghbolagh:2022uij,
  author = "Babaei-Aghbolagh, H. and Velni, Komeil Babaei and Yekta, Davood Mahdavian and Mohammadzadeh, H.",
  title = "{Emergence of non-linear electrodynamic theories from TT{\textasciimacron}-like deformations}",
  eprint = "2202.11156",
  archivePrefix = "arXiv",
  primaryClass = "hep-th",
  reportNumber = "IPM/P-2022/13",
  doi = "10.1016/j.physletb.2022.137079",
  journal = "Phys. Lett. B",
  volume = "829",
  pages = "137079",
  year = "2022"
}

@article{Ferko:2022cix,
  author = "Ferko, Christian and Sfondrini, Alessandro and Smith, Liam and Tartaglino-Mazzucchelli, Gabriele",
  title = "{Root-$T \bar T$ Deformations in Two-Dimensional Quantum Field Theories}",
  eprint = "2206.10515",
  archivePrefix = "arXiv",
  primaryClass = "hep-th",
  doi = "10.1103/PhysRevLett.129.201604",
  journal = "Phys. Rev. Lett.",
  volume = "129",
  number = "20",
  pages = "201604",
  year = "2022"
}

@article{Russo:2025fuc,
  author = "Russo, Jorge G. and Townsend, Paul K.",
  title = "{Simplified self-dual electrodynamics}",
  eprint = "2505.08869",
  archivePrefix = "arXiv",
  primaryClass = "hep-th",
  doi = "10.1007/JHEP10(2025)120",
  journal = "JHEP",
  volume = "10",
  pages = "120",
  year = "2025"
}

@article{Gaillard:1981rj,
  author = "Gaillard, Mary K. and Zumino, Bruno",
  title = "{Duality Rotations for Interacting Fields}",
  reportNumber = "LAPP-TH-37, CERN-TH-3078",
  doi = "10.1016/0550-3213(81)90527-7",
  journal = "Nucl. Phys. B",
  volume = "193",
  pages = "221--244",
  year = "1981"
}

@article{Gibbons:1995cv,
  author = "Gibbons, G. W. and Rasheed, D. A.",
  title = "{Electric - magnetic duality rotations in nonlinear electrodynamics}",
  eprint = "hep-th/9506035",
  archivePrefix = "arXiv",
  doi = "10.1016/0550-3213(95)00409-L",
  journal = "Nucl. Phys. B",
  volume = "454",
  pages = "185--206",
  year = "1995"
}

@article{Russo:2024ptw,
  author = "Russo, Jorge G. and Townsend, Paul K.",
  title = "{Dualities of self-dual nonlinear electrodynamics}",
  eprint = "2407.02577",
  archivePrefix = "arXiv",
  primaryClass = "hep-th",
  doi = "10.1007/JHEP09(2024)107",
  journal = "JHEP",
  volume = "09",
  pages = "107",
  year = "2024"
}

@article{Russo:2024llm,
    author = "Russo, Jorge G. and Townsend, Paul K.",
    title = "{Causal self-dual electrodynamics}",
    eprint = "2401.06707",
    archivePrefix = "arXiv",
    primaryClass = "hep-th",
    doi = "10.1103/PhysRevD.109.105023",
    journal = "Phys. Rev. D",
    volume = "109",
    number = "10",
    pages = "105023",
    year = "2024"
}

@article{Russo:2024xnh,
    author = "Russo, Jorge G. and Townsend, Paul K.",
    title = "{Causality and energy conditions in nonlinear electrodynamics}",
    eprint = "2404.09994",
    archivePrefix = "arXiv",
    primaryClass = "hep-th",
    doi = "10.1007/JHEP06(2024)191",
    journal = "JHEP",
    volume = "06",
    pages = "191",
    year = "2024"
}

@article{Babaei-Aghbolagh:2025hlm,
    author = "Babaei-Aghbolagh, H. and Chen, Bin and He, Song",
    title = "{Integrable sigma models and universal root $T\overline{T }$ deformation via Courant-Hilbert approach}",
    eprint = "2509.17075",
    archivePrefix = "arXiv",
    primaryClass = "hep-th",
    doi = "10.1007/JHEP01(2026)108",
    journal = "JHEP",
    volume = "01",
    pages = "108",
    year = "2026"
}

@article{Sorokin:2021tge,
    author = "Sorokin, Dmitri P.",
    title = "{Introductory Notes on Non-linear Electrodynamics and its Applications}",
    eprint = "2112.12118",
    archivePrefix = "arXiv",
    primaryClass = "hep-th",
    doi = "10.1002/prop.202200092",
    journal = "Fortsch. Phys.",
    volume = "70",
    number = "7-8",
    pages = "2200092",
    year = "2022"
}

@article{Kuzenko:2026hbw,
    author = "Kuzenko, Sergei M. and Ruhl, Jonah",
    title = "{Causal self-dual nonlinear electrodynamics from the Born-Infeld theory}",
    eprint = "2605.06193",
    archivePrefix = "arXiv",
    primaryClass = "hep-th",
    journal = "arXiv e-prints",
    month = "5",
    year = "2026"
}

@article{Murk:2024nled,
    author = "Murk, Sebastian and Soranidis, Ioannis",
    title = "{Light rings and causality for nonsingular ultracompact objects sourced by nonlinear electrodynamics}",
    eprint = "2406.07957",
    archivePrefix = "arXiv",
    primaryClass = "gr-qc",
    doi = "10.1103/PhysRevD.110.044064",
    journal = "Phys. Rev. D",
    volume = "110",
    number = "4",
    pages = "044064",
    year = "2024"
}

@article{Russo:2026bhnled,
    author = "Russo, Jorge G. and Townsend, Paul K.",
    title = "{Black holes and causal nonlinear electrodynamics}",
    eprint = "2601.07789",
    archivePrefix = "arXiv",
    primaryClass = "hep-th",
    journal = "arXiv e-prints",
    month = "1",
    year = "2026"
}

@book{Rychkov:2016iqz,
  author = "Rychkov, Slava",
  title = "{EPFL Lectures on Conformal Field Theory in D>= 3 Dimensions}",
  eprint = "1601.05000",
  archivePrefix = "arXiv",
  primaryClass = "hep-th",
  doi = "10.1007/978-3-319-43626-5",
  publisher = "Springer",
  series = "SpringerBriefs in Physics",
  year = "2017"
}

\clearpage
\section*{End Matter}

The following appendices provide the specialist derivations cited at their
points of use in the Letter.  Appendix A resolves the auxiliary geometry,
response eigenmodes, full Fresnel cones, Hamiltonian causality check, and
scaling identities; Appendix B gives the exact energy conditions and the
integrable sigma-model realization, and Appendix C collects the analyticity
test, conditional unitarity window, and Lambert-function branches.

\subsection*{Appendix A: Auxiliary geometry and response}
For $U,V>0$, introduce a radial invariant and a characteristic rapidity,
\begin{equation}
 U=\frac r2e^{-\chi},\qquad V=\frac r2e^\chi,\qquad
 r=2\sqrt{UV}=|P|.
\label{eq:hyperbolic}
\end{equation}
The auxiliary boost leaves $r$ fixed and translates only the rapidity:
\begin{align}
 \zeta&=r\cosh(\chi-\theta),\qquad
 \mathcal D=-r\sinh(\chi-\theta),\nonumber\\
 \mathcal J&=1+q\theta\tanh(\chi-\theta).
\label{eq:rapidity}
\end{align}
The caustic envelope obeys
\begin{equation}
 \tanh(\chi-\theta)=-\frac1{q\theta},
 \qquad |q\theta|\geq1.
\label{eq:causticlocus}
\end{equation}
Thus the flow is one-dimensional along each hyperbola
$UV=\mathrm{const}$, and multivaluedness requires $|q\theta|\geq1$.

At a caustic, the full Hessian is
\begin{equation}
 \partial_i\partial_j\Lagr_\alpha
 =-\frac{q\theta}{\zeta\mathcal J}
 \begin{pmatrix}e^{2\theta}&1\\[1pt]1&e^{-2\theta}\end{pmatrix}_{ij},
 \qquad (i,j)=(U,V).
\label{eq:hessian}
\end{equation}
It has rank one and diverges only through $1/\mathcal J$.  Its null and
critical directions in the $(U,V)$ field space are
\begin{align}
 \bm n_0&=(e^{-\theta},-e^\theta),\qquad
 \bm n_c=(e^\theta,e^{-\theta}),\nonumber\\
 h_c&=-\frac{q\theta}{\zeta\mathcal J}
 \left(e^{2\theta}+e^{-2\theta}\right).
\label{eq:responsemodes}
\end{align}
Here $h_c$ is the nonzero Hessian eigenvalue.  Only this eigenvalue becomes
singular, so the fold is a one-channel critical phenomenon.  The master
potential gives, on the same real branch,
\begin{equation}
 \Omega_\alpha'(y)=\frac{U}{y^2}+V,\qquad
 \Omega_\alpha''(y)=\frac{\zeta}{y^2}
 \left[\frac{\alpha}{(\alpha-1)\theta}-1\right],
\label{eq:convexity}
\end{equation}
which directly yields the optical indices in Eq.~\eqref{eq:optical}.

For completeness, choose the static background frame with the common
electric-magnetic direction along the third axis.  The full Fresnel
polynomial factorizes as
\begin{equation}
 \mathcal P(k)\propto
 \prod_{\nu=\pm}\left[
 -\omega^2+k_\parallel^2+A_\nu|\mathbf k_\perp|^2\right].
\label{eq:fresnel}
\end{equation}
Each factor is the null condition for an inverse effective metric
$G_\nu^{\mu\rho}\propto
\operatorname{diag}(-1,A_\nu,A_\nu,1)$ in this frame.
For a wave normal making angle $\beta$ with the background direction,
\begin{equation}
 v_{{\rm ph},\nu}^2=\cos^2\beta+A_\nu\sin^2\beta,\quad
 v_{{\rm g},\nu}^2=
 \frac{\cos^2\beta+A_\nu^2\sin^2\beta}
 {\cos^2\beta+A_\nu\sin^2\beta}.
\label{eq:anglevelocities}
\end{equation}
Hence $0<A_\nu\leq1$ is necessary and sufficient for real,
nonsuperluminal characteristics at every angle.  If $A_\nu>1$, every
nonparallel ray is superluminal; if $A_\nu<0$, sufficiently transverse
wave normals are nonhyperbolic.  This proves that the transverse indices in
Fig.~\ref{fig:phasefold} bound the complete angular problem.

The Hamiltonian check can also be made algebraic.  On $U=0$, define
$\ell(V)=\Lagr(0,V)$ and
$\sigma=V[\ell'(V)]^2$.  The Lagrangian and Hamiltonian
Courant--Hilbert functions obey
$\mathfrak h'(\sigma)\ell'(V)=1$.  Using $\ell'=e^{-\theta}$ and
Eq.~\eqref{eq:Vtheta} gives
\begin{equation}
 \mathfrak h'=e^\theta,\qquad
 \mathfrak h''=\frac{q\theta e^\theta}
 {\sigma(1-q\theta)},\qquad
 \mathfrak h'+2\sigma\mathfrak h''
 =e^\theta\frac{1+q\theta}{1-q\theta}.
\label{eq:Hcheck}
\end{equation}
For $q<0$, these expressions establish the violations stated below
Eq.~\eqref{eq:Hcausality}.  They also show that the $\theta>0$ fold
$1+q\theta=0$ coincides with degeneration of the strong-field
hyperbolicity inequality.

The weak-coupling expansion is
\begin{equation}
 \begin{split}
 \Lagr_\alpha={}&V-U-\lambda(U+V)^{1/\alpha}\\
 &+\frac{\lambda^2}{2\alpha^2}(V-U)
 (U+V)^{2/\alpha-2}+O(\lambda^3).
 \end{split}
\label{eq:expansion}
\end{equation}
The exact solution also obeys
\begin{equation}
 \left[
 U\partial_U+V\partial_V+
 \left(1-\frac1\alpha\right)\lambda\partial_\lambda
 \right]\Lagr_\alpha=\Lagr_\alpha,
\label{eq:homogeneity}
\end{equation}
or, for $\alpha\neq1$,
\begin{align}
 \Lagr_\alpha(U,V;\lambda)
 &=\Lambda_\lambda\,
 \mathcal F_\alpha\!\left(
 \frac{U}{\Lambda_\lambda},\frac{V}{\Lambda_\lambda};
 \operatorname{sgn}\lambda\right),\nonumber\\
 \Lambda_\lambda&=|\lambda|^{\frac{\alpha}{\alpha-1}}.
\label{eq:scalingcollapse}
\end{align}
The critical field in Eq.~\eqref{eq:criticalfield} is the exact
dimensionless multiple of this unique field-density scale.

\subsection*{Appendix B: Energy and integrable realization}
For a homogeneous electric field, decompose the pressure into components
parallel and transverse to the field.  The exact stress tensor gives
\begin{align}
 \rho&=\mathcal H=\zeta[1+(\alpha-1)\theta],\qquad
 p_\parallel=-\rho,\nonumber\\
 p_\perp&=\Lagr=\zeta[1-(\alpha-1)\theta].
\label{eq:pressures}
\end{align}
On the relevant branch with $\lambda>0$ and
$0<\theta<\theta_c$, the weak and dominant energy conditions hold up to
the fold:
\begin{equation}
 \rho+p_\perp=2\zeta>0,\qquad
 \rho^2-p_\perp^2=4(\alpha-1)\theta\zeta^2>0.
\label{eq:energyconditions}
\end{equation}
The strong energy condition holds only for
$\theta\leq1/(\alpha-1)$, whereas optical causality fails for every
$\theta>0$.  The fold therefore cannot be diagnosed from positivity or
finiteness of the energy density alone~\cite{Russo:2024xnh}.

For the two-dimensional realization, let $g$ be a group-valued field,
$\partial_\pm$ light-cone derivatives, and $\Tr$ an invariant bilinear form
on the Lie algebra.  With $j_\pm=g^{-1}\partial_\pm g$, define
\begin{align}
 P_1&=-\Tr(j_+j_-),\nonumber\\
 P_2&=\frac12\left[\Tr(j_+j_+)\Tr(j_-j_-)
                    +\Tr(j_+j_-)^2\right],\nonumber\\
 u&=\frac{\sqrt{2P_2-P_1^2}-P_1}{4},\qquad
 v=\frac{\sqrt{2P_2-P_1^2}+P_1}{4}.
\label{eq:uv}
\end{align}
These variables reduce Lax flatness to $\Lagr_u\Lagr_v=-1$.
Both exact representations lead to the same sigma-model action: one may
either replace $(U,V)$ by $(u,v)$ directly in
Eqs.~\eqref{eq:exactL}--\eqref{eq:aux}, or make the same replacement in
Eq.~\eqref{eq:masterpotential} and eliminate $y$ by stationarity.  Expanding
either form about $\theta=0$ (equivalently $y=1$) gives
Eq.~\eqref{eq:expansion}.  Since Eq.~\eqref{eq:uv} implies
\[
 v-u=\frac{P_1}{2},\qquad
 u+v=\frac12\sqrt{2P_2-P_1^2},
\]
the replacement $(U,V)\to(u,v)$ in Eq.~\eqref{eq:expansion} gives
\begin{equation}
 \begin{split}
 \Lagr_\alpha={}&\frac{P_1}{2}
 -\frac{\lambda}{2^{1/\alpha}}
 (2P_2-P_1^2)^{1/(2\alpha)}\\
 &+\frac{\lambda^2P_1}{2^{2/\alpha}\alpha^2}
 (2P_2-P_1^2)^{1/\alpha-1}
 +O(\lambda^3).
 \end{split}
\label{eq:sigmaexpansion}
\end{equation}
For $\alpha=2$ this gives
\begin{equation}
 \Lagr_2=\frac{P_1}{2}
 -\frac{\lambda}{\sqrt2}(2P_2-P_1^2)^{1/4}
 +\frac{\lambda^2P_1}{8\sqrt{2P_2-P_1^2}}
 +O(\lambda^3).
\end{equation}
Here $[\lambda]=1$ and the leading perturbing operator has dimension one.
The interaction is nonanalytic at $P_1=P_2=0$, but the classical Lax
condition is exact away from that locus.  On $P_2=P_1^2$, $P_1>0$, the
auxiliary equation reduces to Eq.~\eqref{eq:Lambertgeneral} with
$V=P_1/2$, and hence
\begin{equation}
 P_{1,c}=2V_c,\qquad
 \frac{\partial^2\Lagr}{\partial P_1^2}
 \propto(P_1-P_{1,c})^{-1/2}.
\label{eq:sigmacritical}
\end{equation}

\subsection*{Appendix C: Analyticity and Lambert branches}
The operator that initiates the flow at the Maxwell point is
\begin{equation}
 \mathcal O_\alpha=(U+V)^{1/\alpha}
 =(S^2+P^2)^{1/(2\alpha)},\qquad
 \Delta_{\mathcal O}=\frac d\alpha.
\label{eq:seedoperator}
\end{equation}
Its four-dimensional weak-field expression is analytic only on the
discrete irrelevant sequence $\alpha=1/(2n)$, $n\in\mathbb N$; every
relevant member is nonanalytic.  If a quantum completion were to promote
$\mathcal O_\alpha$ to a scalar primary of a unitary CFT in $d>2$, the
scalar unitarity bound would require
\begin{equation}
 \alpha\leq\frac{2d}{d-2}.
\label{eq:unitaritywindow}
\end{equation}
Thus in four dimensions even this conditional window ends at $\alpha=4$
~\cite{Rychkov:2016iqz}.  This observation does not establish a quantum
completion or prove inconsistency below the bound; it delimits what such a
completion would have to satisfy.

On the electric axis, define
$q=(1-\alpha)/\alpha$ and
$\kappa=(1-\alpha)\lambda/\alpha^2$.  The auxiliary equation has the exact
solution
\begin{equation}
 \zeta^q=\frac{W(\kappa V^q)}{\kappa},\qquad
 \Lagr_\alpha=\zeta-\frac{\alpha-1}{\alpha}
 \lambda\zeta^{1/\alpha},
\label{eq:Lambertgeneral}
\end{equation}
where $W$ is the Lambert function, $W(z)e^{W(z)}=z$, and its branch is fixed
by $\zeta\to V$ as $\lambda\to0$.
Two cases make the change in analytic structure explicit:
\begin{align}
 \alpha=\tfrac12:\quad
 \zeta&=\frac{W(2\lambda V)}{2\lambda},\nonumber\\[-2pt]
 \Lagr_{1/2}&=V-\lambda V^2+2\lambda^2V^3+\cdots ,
\label{eq:halfexample}\\
 \alpha=2:\quad
 w&=W\!\left(-\frac{\lambda}{4\sqrt V}\right),\nonumber\\[-2pt]
 \Lagr_2&=\frac{\lambda^2(1+2w)}{16w^2},\nonumber\\[-2pt]
 &=V-\lambda\sqrt V+\frac{\lambda^2}{8}+\cdots .
\label{eq:twoexample}
\end{align}
The irrelevant example is analytic at the Maxwell vacuum.  For $\alpha=2$,
Eqs.~\eqref{eq:criticalfield} and~\eqref{eq:foldscaling} give
$V_c=e^2\lambda^2/16$ and $w=-1$.  The first derivative
$\Lagr_V=e^{2w}$ and the energy remain finite, whereas $\Lagr_{VV}$
diverges.  Reversing the flow sign moves the Lambert argument to the
positive real axis and removes this finite-field obstruction, while the
$\sqrt V$ nonanalyticity and the causality obstruction remain.

\end{document}